\documentclass[final]{jfm}
\usepackage{amsfonts,amsbsy,amsgen,amssymb,natbib}
\usepackage[dvips]{graphics}
\usepackage{subfigure}
\usepackage{psfrag}
\usepackage[utf8]{inputenc} 
\usepackage[T1]{fontenc}
\usepackage{amsfonts,amsbsy,amsgen,amssymb}
\usepackage{amsmath}
\usepackage{epsfig}
\usepackage{color}

\renewcommand{\vec}{\mathbf}
\newcommand{\tens}[1]{\boldsymbol{\mathsf{#1}}}
\title[The history force on a small particle in a linearly stratified fluid]{The history force on a small particle in a linearly stratified fluid  }
\author[F. Candelier, R. Mehaddi and O. Vauquelin]
{\textbf{Fabien Candelier$\dagger$, Rabah Mehaddi and Olivier Vauquelin}}
\affiliation{
Aix-Marseille Universit\'e,  CNRS, IUSTI UMR 7343\\
5 rue Enrico Fermi, 13 013 Marseille Cedex 13, France}

\begin{document}
\maketitle
\begin{abstract}
The hydrodynamic force experienced by  a small spherical particle undergoing an arbitrary  time-dependent motion in a weakly density-stratified fluid is investigated theoretically. The study is carried out under the Oberbeck-Boussinesq approximation and  in the limit of small Reynolds and small  P\'eclet numbers. 
The force acting on the particle is obtained by using matched asymptotic expansions. In this approach, the small parameter is given by $a/\ell$, where $a$ is the particle radius and $\ell$ is the stratification length, as defined by \cite{Ardekani10}, which depends on the Brunt-V\"ais\"al\"a frequency, on the fluid kinematic viscosity and on the thermal or the concentration diffusivity (depending on the case considered).  The matching procedure used here, which is based on series expansions of generalized functions, slightly differs from that generally used in similar  problems. In addition to the classical Stokes drag, it is found that the particle experiences a memory force given by two convolution products, one of which involves, as usual, the particle acceleration and the other one, the particle velocity.  Owing to the stratification, the transient behaviour of this memory force, in response to an abrupt motion, consists of an initial fast decrease followed by a damped oscillation with an angular-frequency corresponding to the Brunt-V\"ais\"al\"a frequency.  The perturbation force eventually tends to a constant which provides us with correction terms that should be added to the  Stokes drag to  accurately predict the settling time of a particle  in a diffusive stratified-fluid. 
\end{abstract}

\section{Introduction}

The oceans, or the lakes are good  examples of natural environments in which a density-stratification induced either by a gradient of concentration of a given element in the fluid, or by a gradient of temperature, is often observed. Density-stratified fluid are also widely encountered in industrial processes involving heated fluid or the mixing of fluids of different densities, or again, in fire engineering,  to name but a few.

Today, it is well known that a particle which is settling in a vertically density-stratified fluid experiences a greater resistive force than that which would be measured in a homogeneous fluid, owing to buoyancy effects. Indeed, in such a case, the density gradient modifies the (perturbation) fluid flow produced by the particle, and in particular,  tends to inhibit the vertical motion of the fluid \citep[see for instance][]{ Turner73}. This phenomenon  has been a subject of investigation for many years, owing to the wide range of engineering and environmental applications where the role played by the stratification is of importance \citep[for a review, see the introduction of the article by][and references therein]{Yick09}. However, in the great majority of papers available in the literature, it turns out  that the Reynolds numbers of the particles are relatively high or, at least,  moderate. 

In the creeping flow limit (i.e. small particle Reynolds numbers),  considerably less works exist, despite this problem shows many fundamental aspects, and has also obvious applications in the fields of physics or biology (see for instance MacIntyre, Alldredge \& Gotschalk  1995). By using a method of matched asymptotic expansions similar to that devised by  \cite{Childress64} \citep[see also][]{Saffman65},  \cite{Chadwick74} have investigated the steady force acting on a particle moving horizontally in a non-diffusive stratified fluid. In the same period, they have also investigated the force acting on a particle  settling vertically in a slightly diffusive stratified fluid \citep{Zvirin75}, that is under the assumption that the Péclet number of the particle is not too small.  In these two cases, these authors have shown that, even at vanishingly small Reynolds number, the stratification is responsible for a drag enhancement, owing to the fact that the buoyancy force involved in the equations governing the fluid motion is no longer negligible far from the inclusion. Such a trend has been recovered in the experiments and numerical simulations by \cite{Yick09}, although the agreement between their results and the theory  by \cite{Zvirin75} is not perfect.

In the limit where both the Péclet number and the Reynolds number are small compared to unity, the flow  produced by a settling particle (actually a point force) in a stratified fluid  has been recently investigated by \cite{Ardekani10}. In particular, these authors have exhibited  the existence of a fundamental length scale, characterizing the  fluid stratification, and which is defined by
\begin{equation}
\ell=\left(\frac{\nu \kappa}{N^2}\right)^{1/4}\:,
\label{def_l}
\end{equation}
where 
$$
N = \sqrt{-\frac{g}{\rho_\infty} \frac{\mbox{d}\rho_0}{\mbox{d}z}}
$$
 is the Brunt-V\"ais\"al\"a frequency, $\rho_0$ is the unperturbed fluid density, $\nu$ is the fluid kinematic viscosity and $\kappa$ is the thermal or the concentration diffusivity (depending on the case considered). This length, which reflects the competition of buoyancy, diffusion and viscosity within the fluid,  has been shown to play a significant role in the fluid dynamics when the Péclet number is small compared to unity. However, though the study by \cite{Ardekani10} provides us with an accurate description of the flow produced by a particle moving in a diffusive stratified-fluid, information  concerning the force acting on it is missing, and many questions concerning the dynamics of a particle in a stratified fluid remain open. 

The objectives of the present study are to investigate theoretically the combined effects of the buoyancy and of the unsteadiness on the drag acting on a particle in a stratified fluid, at small but finite Péclet and Reynolds numbers. More specifically, in the same vein as the studies by \cite{Ockendon68,Sano81,Lovalenti93,Lawrence95},  
where the history force  \citep[][]{Boussinesq85,Basset88} has been shown to be modified  by weak convective inertia effects, 
we will focus on how the long-time behaviour of the history force is altered by buoyancy effects. This theoretical analysis will finally enable us to see how  the results by \cite{Zvirin75} can be completed under the assumptions of the present study. 

\section{Governing equations}

In this section, the basic assumptions made in this analysis are detailed, as well as the governing equations for the fluid and density perturbations. Since the force acting on the particle will be carried out by means of matched-asymptotic expansions, we also analyse how these equations can be simplified in a region close to the particle (defined as the inner zone)  or in a region located far from it (outer zone). 

\subsection{General equations}

We consider the arbitrary motion of a small spherical particle of radius $a$ in a vertically density-stratified fluid. The  unperturbed density  is assumed to vary linearly with respect to the vertical coordinate $z$ (measured relative to the initial position of the particle)
$$
\rho_0 = \rho_\infty - \gamma \:z \:,
$$
where $\gamma$ is a positive constant and $\rho_\infty$ is the reference density. Under this condition, the stratification is stable, so that the  unperturbed pressure field is provided by the hydrostatic equation
\begin{equation}
-\boldsymbol{\nabla} p_0 +\rho_0 \:\vec{g} = \vec{0}\:,
\label{eq_statique}
\end{equation}
where $\vec{g}=-g \vec{e}_3$ is the gravity acceleration.  

The density and the pressure variations, with respect to the hydrostatic solution,  induced by the particle when it moves are expected to be weak,  allowing us to assume the Boussinesq-Oberbeck approximation \citep[see for instance][]{Landau89} to be valid. Accordingly, the perturbation fluid motion equations, written in a frame of reference moving with the particle,  read as
\begin{equation}
\boldsymbol{\vec{\nabla}}\cdot \vec{w} = 0\:,
\label{incomp}
\end{equation}
\begin{eqnarray}
\rho \left(\frac{\partial \vec{w}}{\partial t}  - (\vec{u}\cdot \boldsymbol{\nabla}) \vec{w}+ (\vec{w}\cdot \boldsymbol{\nabla}) \vec{w}
\right)  = - \boldsymbol{\nabla} p
+ \rho \vec{g}+ \rho \nu \boldsymbol{\triangle}\vec{w}\:,\label{eq_mvt_1}\\
\vec{w} = \vec{u} \:,\quad  r=a \quad \mbox{and} \quad 
\vec{w} \to \vec{0} \:,\quad r\to \infty\:, \label{eq_cl1}
\end{eqnarray}
where $\rho$ and $p$ correspond to the actual density and pressure of the fluid,  and $\vec{u}$ and $\vec{w}$ denote respectively the particle and the perturbation fluid velocities. 
By  introducing the following decompositions in (\ref{eq_mvt_1})
$$
\rho = \rho_0 + \rho' =  \rho_\infty\left( 1- \frac{\gamma z}{\rho_\infty} + \frac{\rho'}{\rho_\infty} \right) \quad \mbox{ and } \quad p = p_0 + p'\:,
$$
where $\rho'$ and $p'$ are  the perturbations of the fluid density  and of the  pressure, and according to (\ref{eq_statique}), (\ref{eq_mvt_1}) can be recast as
$$
\rho_\infty\left(\frac{\partial \vec{w}}{\partial t}  - (\vec{u}\cdot \boldsymbol{\nabla}) \vec{w}+ (\vec{w}\cdot \boldsymbol{\nabla}) \vec{w}
\right)  = -\frac{\boldsymbol{\nabla} p'}{\left( 1- \displaystyle{\frac{\gamma z}{\rho_\infty}}+ \frac{\rho'}{\rho_\infty} \right)}
+ \frac{\rho'\vec{g}}{\left( 1-  \displaystyle{\frac{\gamma z}{\rho_\infty} + \frac{\rho'}{\rho_\infty} }\right)}+ \rho_\infty \nu \boldsymbol{\triangle}\vec{w}\:.
$$
Denoting by $d$ the  distance from the sphere defining the outer region   (which is unknown for the moment), and by further assuming that the condition
\begin{equation}
\frac{\gamma d}{\rho_\infty} \ll1 
\label{condition_B}
\end{equation}
is fulfilled, it turns out that  (\ref{eq_mvt_1}) can be approximated here by
\begin{equation}
\rho_\infty\left(\frac{\partial \vec{w}}{\partial t}  - (\vec{u}\cdot \boldsymbol{\nabla}) \vec{w}+ (\vec{w}\cdot \boldsymbol{\nabla}) \vec{w}
\right)  = - \boldsymbol{\nabla} p'
+ \rho'\vec{g}+ \rho_\infty \nu \boldsymbol{\triangle}\vec{w}\:,\label{eq_mvt_1_2}
\end{equation}
where quadratically small terms of the form $\left({\gamma d}/{\rho_\infty}\:,\:{\rho'}/{\rho_\infty}\right) \times \left(\nabla p'\:,\:\rho' g\right)$  are neglected.

The density $\rho'$ involved in (\ref{eq_mvt_1_2}) is assumed to be governed by a (classical) diffusion-advection equation which, in the present case, reads as
\begin{equation}
\frac{\partial \rho'}{\partial t} - \gamma \:\vec{w}\cdot \vec{e}_3 - (\vec{u}\cdot \boldsymbol{\nabla}) \rho'  + (\vec{w}\cdot \boldsymbol{\nabla}) \rho' = \kappa \triangle \rho'\:, 
\label{eq_rho}
\end{equation}
with
\begin{equation}
\frac{\partial \rho'}{\partial r} \Big|_{r=a}  = - \frac{\partial \rho_0}{\partial r}\Big|_{r=a}\:,\quad  r=a \quad \mbox{and} \quad 
\rho' \to  0\:,\quad r\to \infty\:.
\label{eq_cl2}
\end{equation}
Note that for the sake of simplicity, in the boundary condition (\ref{eq_cl2}), the particle has been assumed to be either adiabatic or impermeable according to whether the stratification is  induced by  a gradient of temperature or by  a gradient of concentration of a given element in the fluid.

In order to normalize  (\ref{eq_mvt_1_2}) and (\ref{eq_rho}), let us notice now that, at the vicinity of the sphere (i.e.  $r \sim a$), the boundary conditions  (\ref{eq_cl1}) and (\ref{eq_cl2}) suggest that the perturbation fluid velocity scales as the  particle velocity  and that the  density perturbation scales as $\rho' \sim \gamma a$. Thus by using  $\gamma a$ in order to scale the density, $a$ for lengths,  $\rho_\infty \nu u / a$ for the pressure, and by using an arbitrary (typical) velocity $u$ for the particle and fluid velocities,  we are led to 
\begin{equation}
\boldsymbol{\vec{\nabla}}\cdot \vec{w} = 0\:,
\label{incomp}
\end{equation}
\begin{equation}
\frac{a^2}{\nu \tau} \frac{\partial \vec{w}}{\partial t}  +\mbox{Re}\Big(- (\vec{u}\cdot \boldsymbol{\nabla}) \vec{w} + (\vec{w}\cdot \boldsymbol{\nabla}) \vec{w}
\Big) = - \boldsymbol{\nabla} p'
- \frac{a^3\gamma g}{\rho_\infty \nu u  } \:\rho'\:\vec{e}_3 + \boldsymbol{\triangle}\vec{w}\:,\label{eq_mvt_1_adim1}
\end{equation}
\begin{equation}
\vec{w} = \vec{u} \:,\quad  r=1 \quad \mbox{and} \quad 
\vec{w} \to \vec{0} \:,\quad r\to \infty\:,
\end{equation}
and to
\begin{eqnarray}
\frac{a^2}{\kappa \tau} \frac{\partial \rho'}{\partial t} + \mbox{Pe}\Big(- \:\vec{w}\cdot \vec{e}_3  -(\vec{u}\cdot \boldsymbol{\nabla}) \rho'+ (\vec{w}\cdot \boldsymbol{\nabla}) \rho' \Big) = \triangle \rho'\:, 
\label{eq_rho_2}\\
\frac{\partial \rho'}{\partial r} \Big|_{r=1}  = \cos(\theta)\:,\quad  r=1 \quad \mbox{and} \quad 
\rho' \to  0\:,\quad r\to \infty\:.
\label{eq_cl2_2}
\end{eqnarray}
In these equations, $\tau$ is a characteristic time that will be specified shortly, $\mathrm{Re}=au/\nu$ and $\mathrm{Pe}=au/\kappa$ are the particle Reylnolds and Péclet numbers, and $\theta$ is the angle between the radial unit vector $\vec{e}_r$ and the vertical direction $\vec{e}_3$.  Note  that for the sake of simplicity, notations have not changed, though they are now related to non-dimensional variables.

At this stage, it is worth mentioning that in the study by \cite{Zvirin75}, the force corrections obtained for a particle settling in a slightly diffusing stratified fluid has been found to depend  on a  so-called stratification number \citep[redefined recently by][as a viscous Richardson number]{Yick09}
$$
\mathrm{Ri} =\frac{a^3\gamma g}{\rho_\infty \nu u  } =  \frac{a^3\:N^2}{u\:\nu}\:,
$$
which naturally appears in (\ref{eq_mvt_1_adim1}). 
In the present study, it turns out to be more convenient to introduce, instead of this Richardson  number, an alternative non-dimensional number  based on the ratio between the particle radius $a$,  and the stratification length $\ell$ defined by \cite{Ardekani10} (see \ref{def_l}), and which is linked to the Richardson number according to the relation
\begin{equation}
\mathrm{Ri} = \frac{1}{\mathrm{Pe}}\left(\frac{a}{\ell}\right)^4 \:.
\label{def_Ri}
\end{equation}
After that, for reasons of mathematical convenience, we also introduce in the equations,   a stretched density perturbation 
$$
\rho' =  \mathrm{Pe} \:\tilde{\rho} \:, 
$$
so that  (\ref{eq_mvt_1_adim1}) can be re-written as 
\begin{equation}
\frac{a^2}{\nu \tau} \frac{\partial \vec{w}}{\partial t}  +\mbox{Re}\Big(- (\vec{u}\cdot \boldsymbol{\nabla}) \vec{w} + (\vec{w}\cdot \boldsymbol{\nabla}) \vec{w}
\Big) = - \boldsymbol{\nabla} p'
-\left(\frac{a}{\ell}\right)^4\:\tilde{\rho}\:\vec{e}_3 + \boldsymbol{\triangle}\vec{w}\:,\label{eq_mvt_1_adim2}
\end{equation}
and equations (\ref{eq_rho_2}) and (\ref{eq_cl2_2}) become
\begin{eqnarray}
\frac{a^2}{\kappa \tau} \frac{\partial \tilde{\rho}}{\partial t} - \:\vec{w}\cdot \vec{e}_3 + \mbox{Pe}\Big( -(\vec{u}\cdot \boldsymbol{\nabla}) \tilde{\rho}+ (\vec{w}\cdot \boldsymbol{\nabla}) \tilde{\rho} \Big) = \triangle \tilde{\rho}\:, 
\label{eq_rho_2_2}\\
\frac{\partial \tilde{\rho}}{\partial r} \Big|_{r=1}  = \frac{\cos(\theta)}{\mathrm{Pe}}\:,\quad  r=1 \quad \mbox{and} \quad 
\tilde{\rho} \to  0\:,\quad r\to \infty\:.
\label{eq_clrho_2}
\end{eqnarray}

Equations (\ref{eq_mvt_1_adim2}) and (\ref{eq_rho_2_2}), together with their related boundary conditions, provide us a set of equations that  will be further simplified according to additional assumptions.

\subsection{The inner equations}

As stated in the introduction, the velocity of the particle is assumed to be small enough to ensure that both the P\'eclet number and the Reynolds number remain small compared to unity.  If we further consider that the ratio $a/\ell$, as well as the unsteady terms in (\ref{eq_mvt_1_adim2}) and (\ref{eq_rho_2_2}) are also small parameters, then at the vicinity of the particle, and at leading-order, the governing  equations degenerate into 
\begin{equation}
0 = - \boldsymbol{\nabla} p'
 + \boldsymbol{\triangle}\vec{w}\quad ; \quad \vec{w} = \vec{u} \:,\:  r=1 
 \label{eq_mvt_1_adim3}
\end{equation}
and
\begin{equation}
 - \:\vec{w}\cdot \vec{e}_3 = \triangle \tilde{\rho}\:\quad ; \quad 
\frac{\partial \tilde{\rho}}{\partial r} \Big|_{r=1}  = \frac{\cos(\theta)}{\mathrm{Pe}}\:,\:  r=1 \:.
\label{eq_cl2_2_3}
\end{equation}
Let us denote by $\vec{w}_0$, $p'_0$ and $\tilde{\rho}_0$ the solutions of (\ref{eq_mvt_1_adim3}) and (\ref{eq_cl2_2_3}). 
As expected, it is seen that  $\vec{w}_0$, $p'_0$ simply correspond  to the classical Stokes solution \citep[see for instance][]{Happel83}. As for the density perturbation, we shall not give here its general analytical expression since it is not really needed in the following. Let us simply mention that in a general case, the forced part of the solution induced by the presence of the fluid velocity component in (\ref{eq_cl2_2_3}) actually contains a linearly increasing term, so that in the inner region, the density scales as
\begin{equation}
\tilde{\rho}_0  \sim \mbox{max}\left( \frac{1}{\mathrm{Pe}\:r^2}\:,\:r \right) \:. 
\label{eq_rho_0}
\end{equation}
This is indicative of the fact that this problem requires matched-asymptotic expansion to be fully solved, since the increasing part of the solution is not compatible with the vanishing of the density as suggested by the boundary condition (\ref{eq_clrho_2}). 
In other words, the inner solution has to be matched to an outer solution, valid in a region far from the particle. 

\subsection{The outer equations}

For the moment, the only thing that we can say about the far-field region is that it is defined by $r\gg 1$. 
To simplify (\ref{eq_mvt_1_adim2}) and (\ref{eq_rho_2_2}) in this region,  the order of magnitude of  the convective terms  are now compared with the diffusive terms. Note that for simplicity, such comparisons will be made  in the steady limit, which corresponds to the case where the penetration depth of the disturbance flow is maximal, and thus,  conclusions that will be drawn ought to be valid also in the unsteady case.   To do so, we  use  the fact that the leading order of the fluid velocity found in the inner zone decreases as $O(1/r)$ (since it 
corresponds to a Stokes flow) so that  we are led to
\begin{equation}
\mbox{Re}  (\vec{u} \cdot \nabla) \vec{w}   \sim O\left(\frac{\mbox{Re}}{r^2}\right) \:,\quad \mbox{Re}  (\vec{w} \cdot \nabla) \vec{w} \sim O\left(\frac{\mbox{Re}}{r^3}\right)
\quad \mbox{and} \quad 
\quad \triangle \vec{w} \sim O\left( \frac{1}{r^3}\right)\:.
\label{eq_convec_w}
\end{equation}
As usual in this kind of problem,  it is observed that the former convective terms balance the viscous terms at a distance given by
$r \sim 1/\mbox{Re} $ (Oseen's length), whereas the second one is always smaller than it.  Also, according to 
(\ref{eq_rho_0}) and by assuming that
in the outer zone, $r \gg 1/(\mathrm{Pe} \:r^2)$, 
 it is seen that the buoyancy force balances the  diffusive terms at a distance given by  
$$
r   \sim \frac{\ell}{a} \:.
$$

In a similar way,  in  (\ref{eq_rho_2_2}), it may be shown that 
\begin{equation}
\mbox{Pe}(\vec{u}\cdot \boldsymbol{\nabla}) \tilde{\rho}  \sim O \left(\mbox{Pe}\right)  \:,  \quad  \mbox{Pe} (\vec{w}\cdot \boldsymbol{\nabla}) \tilde{\rho} \sim O\left(\frac{\mbox{Pe}}{r}\right)\: \quad \mbox{and} \quad \triangle \tilde{\rho} \sim O\left(\frac{1}{r}\right)
\label{eq_convec_rho}
\end{equation}
so that  the former convective term  balances the diffusive term at a distance given by $r \sim 1/\mbox{Pe} $, whereas the second one is always negligible.

According to the results provided by (\ref{eq_convec_w}) and (\ref{eq_convec_rho}), we shall assume in the following that
\begin{equation}
\frac{\ell}{a}   \ll \mbox{min}\left(\frac{1}{\mbox{Re}}\:,\:\frac{1}{\mbox{Pe}}\:,\:\mbox{Pe} \left(\frac{\ell}{a}\right)^4\right) \:.
\label{condition2}
\end{equation}
As result, in the  region defined by $r \sim \ell/a$, both the buoyancy force involved in  (\ref{eq_mvt_1_adim2})  and the convective term $\vec{w} \cdot \vec{e}_3$ in  (\ref{eq_rho_2_2})  have  to be kept whereas all the  other convective terms may  be neglected. The distance $\ell$  therefore defines the outer region of this problem (i.e. $d \sim \ell$), and consequently, a natural choice for  the time scale $\tau$ is to set
$$
\tau = \frac{\ell^2}{\nu}\:,
$$ 
which corresponds to the typical time the vorticity generated by the particle displacement takes to diffuse to the stratification
length $\ell$. Thus, in the far-field  region the equations which are to be considered, in order to  determine the force acting on the particle,  read as
 \begin{equation}
\boldsymbol{\vec{\nabla}}\cdot \vec{w} = 0\:,
\label{incomp}
\end{equation}
\begin{eqnarray}
\label{eq_mvt_3}
 \left(\frac{a}{\ell}\right)^2 \frac{\partial \vec{w}}{\partial t}  
= - \boldsymbol{\nabla} p'
- \left(\frac{a}{\ell}\right)^4 \:\tilde{\rho}\:\vec{e}_3 +  \boldsymbol{\triangle}\vec{w}\:, \label{eq_phys_w}
\end{eqnarray}
\begin{eqnarray}
 \left(\frac{a}{\ell}\right)^2 \:\mbox{Pr} \:\frac{\partial \tilde{\rho}}{\partial t} - \vec{w}\cdot \vec{e}_3  = \triangle\tilde{\rho}\:,\label{eq_phys_rho}
\label{cl2}
\end{eqnarray}
where $\mbox{Pr} = \nu/\kappa$ is the Prandtl number (or the Schmidt number).

\section{Force acting the particle}

By using the results of  the previous section,  the force acting on the particle can now be carried out by means of matched asymptotic expansions, and to do so we introduce the following small parameter 
$\epsilon = {a}/{\ell}$. 

Following the method devised by \cite{Childress64}, in the inner region (defined by $r \sim 1$),  the velocity and the pressure are sought  in the form 
\begin{equation}
\vec{w} = \vec{w}_0 + \epsilon \: \vec{w}_1 + O\left(\epsilon^2\right)\:, \quad \mbox{and}  \quad p' = p_0'+ \epsilon \: p_1' + O\left(\epsilon^2\right)\:.
\label{inner_expansion1}
\end{equation}
In the far-field region, i.e. $r \sim \ell/a \gg  1$,  the sphere is substituted by a point force, whose strength is set equal to the Stokes drag (but with a minus sign), thus yielding the following equations
\begin{equation}
\boldsymbol{\vec{\nabla}}\cdot \vec{w} = 0\:,
\label{eq1}
\end{equation}
\begin{equation}
\epsilon^2\: \frac{\partial \vec{w}}{\partial t}  = - \boldsymbol{\nabla} p'
- \epsilon^4\:\tilde{\rho} \:\vec{e}_3 +\boldsymbol{\triangle}\vec{w} + 6\pi\vec{u}\:\delta(\vec{r})\:,
\label{eq2}
\end{equation}
\begin{equation}
\epsilon^2 \:\mbox{Pr}\:\frac{\partial \tilde{\rho}}{\partial t} - \vec{w}\cdot \vec{e}_3 =  \triangle \tilde{\rho}\:,
\label{eq3}
\end{equation}
where $\delta(\vec{r})$ denotes the Dirac delta function. Note that the absence of a Dirac source term in (\ref{eq3}) stems from the fact that the integration of ${\partial \rho'}/{\partial r}$ over the particle surface is zero, according to (\ref{eq_clrho_2}). 

Let us now define the subsequent temporal Fourier transform and spatial Fourier transform (respectively denoted by $\mathcal{F}_t$ and   $\mathcal{F}$)
\begin{equation}
\vec{\hat{w}}(\vec{k},\:\omega) = \mathcal{F}_t(\mathcal{F}(\vec{w}(\vec{x},\:t))) = \int_{\mathbb{R}}  \int_{\mathbb{R}^3} \vec{w}(\vec{x},\:t) \exp(-i\:\vec{k}\cdot \vec{x} + i \:\omega \:t)\:\mbox{d} \vec{x}\:\mbox{d}t
\label{def_Fourier}
\end{equation}
where $i^2=-1$, and the inverse Fourier transforms
$$
\vec{w}(\vec{x},\:t) = \frac{1}{(2\pi)^4} \int_{\mathbb{R}}  \int_{\mathbb{R}^3}\vec{\hat{w}}(\vec{k},\:\omega) \exp(i\:\vec{k}\cdot \vec{x} - i\:\omega \:t)\:\mbox{d} \vec{k}  \:\mbox{d}\omega\:.
$$
Applying these Fourier transforms to equations (\ref{eq1}) to (\ref{eq3})  yields an algebraic system 
\begin{equation}
\vec{k}\cdot \vec{\hat{w}} = 0\:,
\label{eq1_fourier}
\end{equation}
\begin{equation}
\epsilon^2 \:i \omega \vec{\hat{w}} =  i \vec{k} \hat{p}'
+ \epsilon^4\:\hat{\tilde{\rho}}\:\vec{e}_3 +k^2\vec{\hat{w}} - 6\pi\vec{\hat{u}} \:,
\label{eq2_fourier}
\end{equation}
\begin{equation}
\epsilon^2\:\mbox{Pr} \: i \omega \hat{\tilde{\rho}}+ \hat{w}_3 =   k^2 \hat{\tilde{\rho}}\:,
\label{eq3_fourier}
\end{equation}
which is to be solved. After a little algebra, the solution of these equations may be written in the form
\begin{equation}
\vec{\hat{w}} = \frac{6\pi}{k^2-i\:\epsilon^2 \omega} \:\tens{G} \cdot \left( \vec{\hat{u}} -\left(\frac{\vec{\hat{u}}\cdot{\vec{k}}}{k^2} \right)\vec{k} \right)\:,
\label{eq_solution_far}
\end{equation}
where 
\begin{equation}
\tens{G} = 
\left(
\begin{array}{ccc}
1\quad &\quad 0 \quad & \displaystyle{\frac{\epsilon^4 k_3 k_1}{(k^2-i\:\epsilon^2 \mbox{Pr}\:\omega) (k^2-i\:\epsilon^2 \:\omega) k^2 + \epsilon^4(k_1^2+k_2^2)} }\\
 & & \\
0 &1 & \displaystyle{\frac{\epsilon^4 k_3 k_2}{(k^2-i\:\epsilon^2 \mbox{Pr}\:\omega) ( k^2-i\:\epsilon^2 \:\omega) k^2 + \epsilon^4(k_1^2+k_2^2)} }\\
& & \\
0 & 0 & \displaystyle{\frac{k^2(k^2- i \:\epsilon^2\mbox{Pr}\:\omega)(k^2-i\:\epsilon^2 \omega)}{(k^2-i\:\epsilon^2 \mbox{Pr}\:\omega) ( k^2-i\:\epsilon^2 \:\omega) k^2 + \epsilon^4(k_1^2+k_2^2)} }\\
\end{array}
\right) \:.
\end{equation}

In order to determine the force acting on the particle, the last step of the method consists in performing the matching between the inner solution and the outer solution. In this problem, it turns out that  the matching procedure which is usually used in these kind of investigations (see Childress 1964, or  Saffman 1965)  provides us with integral terms  which  cannot be evaluated analytically. In order to overcome such a difficulty, we have been led to develop an alternative matching procedure, described in more details in a companion paper (see Candelier, Mehaddi \& Vauquelin 2013). 
This alternative method is based on the fact that the solution (\ref{eq_solution_far}) of the outer problem is to be interpreted as a generalized function (i.e. a distribution) instead of a classical function. Accordingly, by considering the fact that  the parameter $\epsilon$ is small compared to unity, $\vec{\hat{w}}$ can be expanded, with respect to this parameter, in the form
\begin{equation}
\vec{\hat{w}}  = \vec{\hat{\mathcal{T}}}_0 + \epsilon \vec{\hat{\mathcal{T}}}_1 + \epsilon^2 \: \vec{\hat{\mathcal{T}}}_2  + \ldots +  \epsilon^n \: \vec{\hat{\mathcal{T}}}_n 
\quad \mbox{where, by definition,} \quad 
 \vec{\hat{\mathcal{T}}}_n  = \frac{1}{n! }\lim_{\epsilon \to 0} \frac{\mbox{d}^n\vec{\hat{w}}  }{\mbox{d} \epsilon^n}\:.
  \label{series}
 \end{equation}
The main difference between a series expansion performed in the sense of  classical functions and that performed in the sense of generalized functions is that some terms which are found to be zero in the limit when $\epsilon$ tends to zero in the former case may  tend to the delta distribution (or its derivatives) in the second case.  In particular, in the problem we are interested in, it is readily found that the first term of the series, i.e. $\vec{\hat{\mathcal{T}}}_0$, simply corresponds to the Fourier transform of a Stokeslet. In the matching zone (i.e. $r\sim 1/\epsilon$),  the inverse (spatial) Fourier transform $\mathcal{F}^{-1}\left(\vec{\hat{\mathcal{T}}}_0\right)$ therefore matches  the leading order term $\vec{w}_0$ of the inner expansion  (\ref{inner_expansion1}). 

The second term is found to be of the form
$$
\hat{\mathcal{T}}_1  = \lim_{\epsilon\to 0} \frac{1}{\epsilon^3} \:\vec{f}\left(\frac{\vec{k}}{\epsilon},\:\omega\right)\:,
$$
which means that, in terms of generalized function 
$$
\hat{\mathcal{T}}_1  = -\vec{h}(\omega)  \:\delta(\vec{k})  \quad \mbox{where} \quad   -\vec{h}(\omega) = \int_{\mathbb{R}^3} \frac{\mbox{d} \vec{\hat{w}}  }{\mbox{d}\epsilon} \Big|_{\epsilon=1} \: \mbox{d} \vec{k}\:,
$$
and where the minus sign has been added arbitrarily for convenience. 

According to this result, and by using the fact that the inverse spatial Fourier transform of $\delta(\vec{k})$ is given by $1/(8\pi)^3$
it can be deduced that in the matching region, which is characterized by $r\sim 1/\epsilon$, 
$$
\lim_{r\to 1/\epsilon} \vec{w}_1  \sim 
-\mathcal{F}_t^{-1}\left(\frac{\vec{h}(\omega)}{8\pi^3}\right)  \:.
$$
Similarly as in the classical method, the perturbation term $\vec{w}_1$ of the inner expansion (\ref{inner_expansion1}) therefore matches a time-dependent uniform flow.
As a result, this outer uniform flow produces a drag enhancement which corresponds, at leading order, to an additional  Stokes drag based on this uniform velocity. Hence,  the force induced by the perturbation flow on the particle reads as
\begin{equation}
\vec{f}= -6\:\pi \left(\vec{u} + \epsilon \:\mathcal{F}_t^{-1}\left(\frac{\vec{h}(\omega)}{8\pi^3} \right) \right)\:.
\end{equation}

In what follows, the expression of  $\vec{h}(\omega)$ is detailed and its asymptotic behaviours, with respect to $\omega$, are discussed.

\subsection{Force correction in the frequency domain}
\label{section_force_freq}

In order to determine $\vec{h}(\omega)$, we have used a set of spherical coordinates $k_1 = k \sin(\theta) \cos(\phi)$, $k_2 = k \sin(\theta) \sin(\phi)$, and $k_3 = k \cos(\theta)$, where $k=|\vec{k}|$, $\phi \in [0\:,\:2\pi]$ and $\theta \in [0\:,\:\pi]$.   Integration with respect to $k$ and $\phi$ provides us with the following result 
\begin{equation}
\frac{\vec{h}(\omega)}{8\pi^3}    = h_\bot(\omega) \Big( \hat{u}_1(\omega)\:\vec{e}_1 + \hat{u}_2(\omega)\:\vec{e}_2 \Big) 
+ h_\parallel(\omega)\: \hat{u}_3(\omega)\:\vec{e}_3\:,
\label{eq_res_1}
\end{equation}
where 
\begin{equation}
%\begin{split}
 h_\bot(\omega)  =  I_1(\omega) + \overline{I_1}(-\omega) - \omega^2 \left( I_2(\omega)+\overline{I_2}(-\omega) \right)  
 \:  -i\: \omega \left( I_3(\omega) +\overline{I_3}(-\omega) +\frac{3\sqrt{2}(1+i)}{8\sqrt{\omega}} \right)   
 \label{eq_h_bot}
% \end{split}
\end{equation}
and 
\begin{equation} 
 h_\parallel =   I_4(\omega) +\overline{I_4}(-\omega)- \omega^2 \left(I_5(\omega)+\overline{I_5}(-\omega)\right)\\  
 -i\: \omega
\left( I_6(\omega) +\overline{I_6}(-\omega)\right)\:.
\label{eq_h}
\end{equation}
In these equations,  $I_n$ ($n=1\ldots 6$) are integral terms of the form
\begin{equation}
I_n=\displaystyle{
\int_0^\pi \displaystyle{
\frac{ \frac{3}{8}(1-i) f_n(\theta)}{
\sqrt{\omega^2(\mbox{Pr}-1)^2+4\sin(\theta)^2} \sqrt{\omega (\mbox{Pr}+1)+\sqrt{\omega^2(\mbox{Pr}-1)^2+4\sin(\theta)^2}}}} \mbox{d}\theta}
\label{eq_res2}
\end{equation}
in which 
\begin{eqnarray}
f_1=\sin(\theta)^3\cos(\theta)^2 \:  ,  \: f_2=\frac{1}{2} \sin(\theta)\cos(\theta)^2(\mbox{Pr}-1)\:,\\
f_3=\frac{i}{2} \sin(\theta)\cos(\theta)^2\sqrt{\omega^2(\mbox{Pr}-1)^2+4\sin(\theta)^2}\:,\:f_4 = 2 \sin(\theta)^5\:, \\
f_5=\sin(\theta)^3(\mbox{Pr}-1)\:, \:f_6=i\:\sin(\theta)^3\sqrt{\omega^2(\mbox{Pr}-1)^2+4\sin(\theta)^2}\:,\label{eq_res3}
 \end{eqnarray}
and where the terms $\overline{I_n}$ denote their complex conjugates. 
Except for the case $\mathrm{Pr}=1$, for which solutions can be drawn in terms of elliptic integrals, integration with respect to $\theta$ cannot be performed analytically,  and (\ref{eq_res_1}) has to be  evaluated numerically. 
To go further,  let us detail  the asymptotic behaviours of (\ref{eq_h_bot}) and (\ref{eq_h}) which can be drawn analytically in the cases $\omega \gg 1$ and $\omega \ll 1$. 

In the case of high frequency, and up to $O \left( \left({1}/{\omega}\right)^{7/2} \right)$  we obtain
\begin{align*}
I_1(\omega) + \overline{I_1}(-\omega)  &\sim -\frac{\sqrt{2}(1-i)}{20 ( \mathrm{Pr} + \sqrt{\mathrm{Pr}} ) } \left(\frac{1}{\omega}\right)^{3/2} \:, \\
      - \omega^2 \left( I_2(\omega)+\overline{I_2}(-\omega) \right) &\sim \frac{\sqrt{2}(1-i)}{16} \frac{\sqrt{\mathrm{Pr}}-1}{\sqrt{\mathrm{Pr}}} \sqrt{\omega}+\frac{\sqrt{2}(1-i)}{80}A(\mathrm{Pr}) \left(\frac{1}{\omega}\right)^{3/2} \:,\\
      \begin{split}
-i\: \omega \left( I_3(\omega) +\overline{I_3}(-\omega) +\frac{3\sqrt{2}(1+i)}{8\sqrt{\omega}} \right)  &  \sim  \left(\frac{\sqrt{2}(1-i)}{16} \frac{\sqrt{\mathrm{Pr}}+1}{\sqrt{\mathrm{Pr}}} +
\frac{3\sqrt{2}(1-i)}{8}  \right)\sqrt{\omega}  \\
& \quad + \frac{\sqrt{2}(1-i)}{80} B(\mathrm{Pr}) \left(\frac{1}{\omega}\right)^{3/2}\:,
\end{split}
\end{align*}
where
$$
A(\mathrm{Pr}) = \frac{\mathrm{Pr}^{5/2}-5\mathrm{Pr}^{3/2}+5\mathrm{Pr}-1}{\mathrm{Pr}^{3/2} (\mathrm{Pr}^2-2 \mathrm{Pr}+1)}
\quad \mbox{ and }  \quad
B(\mathrm{Pr}) =  \frac{\mathrm{Pr}+\sqrt{\mathrm{Pr}}+1}{\mathrm{Pr}^2+\mathrm{Pr}^{3/2}}\:,
$$
and 
\begin{align*}
I_4(\omega) +\overline{I_4}(-\omega)  & \sim  - \frac{2 \sqrt{2}(1-i)}{5 ( \mathrm{Pr} + \sqrt{\mathrm{Pr}} ) } \left(\frac{1}{\omega}\right)^{3/2} \:,\\
- \omega^2 \left( I_5(\omega)+\overline{I_5}(-\omega)\right) & \sim  \frac{\sqrt{2}(1-i)}{4}  \frac{\sqrt{\mathrm{Pr}}-1}{\sqrt{\mathrm{Pr}}} \sqrt{\omega}  + \frac{\sqrt{2}(1-i)}{10} A(\mathrm{Pr})\:\left(\frac{1}{\omega}\right)^{3/2}\:,\\
-i\: \omega
\left(I_6(\omega) +\overline{I_6}(-\omega)\right) & \sim  \frac{\sqrt{2}(1-i)}{4}  \frac{\sqrt{\mathrm{Pr}}+1}{\sqrt{\mathrm{Pr}}} \sqrt{\omega} +\frac{\sqrt{2} (1-i)}{10} B(\mathrm{Pr}) \left(\frac{1}{\omega}\right)^{3/2}\:.\\
\end{align*}
From these first asymptotic results, it is observed that the dependence of $ \vec{h}(\omega)$ on the Prandtl number vanishes for the high frequency, since 
$$
\frac{\vec{h}(\omega)}{(2\pi)^3} \sim  \frac{\sqrt{2}}{2} (1-i) \sqrt{\omega} \:\hat{\vec{u}}(\omega) + O\left(\left(\frac{1}{\omega}\right)^3\right)\:.
$$  
Such an expression is similar to that of the classical Boussinesq-Basset history force written in the frequency domain (see for instance Landau \& Lifchitz 1989).
This result is physically sound because for the high frequency, the penetration depth of the disturbance, which scales as $\sqrt{\nu/\omega}$, becomes small compared to $\ell$, so that the effects of the stratification vanish.

On the other hand, that is in the quasi-steady limit (i.e. $\omega \to 0$), the dependence on the Prandtl number also vanishes and up to $O(\omega^2)$ we obtain 
\begin{align*}
I_1(\omega) + \overline{I_1}(-\omega) & \sim \frac{1}{14}\mbox{E}_K\left(\frac{\sqrt{2}}{2}\right)-\frac{3i\:C}{40}  (\mathrm{Pr}+1)\omega \:,\\
- \omega^2 \left\{ I_2(\omega)+\overline{I_2}(-\omega) \right\} & \sim  0\:,\\
-i\: \omega \left\{ I_3(\omega) +\overline{I_3}(-\omega) +\frac{3\sqrt{2}(1+i)}{8\sqrt{\omega}} \right\}  & \sim  \frac{3\sqrt{2}(1-i)}{8}\sqrt{\omega} -\frac{3C}{10} \omega\:,\\
\end{align*}
and 
\begin{align*}
I_4(\omega) + \overline{I_4}(-\omega) & \sim \frac{5}{14}\mbox{E}_K\left(\frac{\sqrt{2}}{2}\right)-\frac{9i C}{40}  (\mathrm{Pr}+1)
\omega \:,\\
- \omega^2 \left\{ I_5(\omega)+\overline{I_5}(-\omega) \right\}  & \sim  0\:,\\
-i\: \omega
\left( I_6(\omega) +\overline{I_6}(-\omega)\right) & \sim -\frac{9C}{10}  \omega \:,\\
\end{align*}
where   $\mbox{E}_k(\:\cdot\:)$ and $\mbox{E}_E(\:\cdot\:)$ are  two complete elliptic integrals, respectively, of the first and of the second kind (see Abramovitz \& Stegun 1965), and $C=\left( \mbox{E}_K\left({\sqrt{2}}/{2}\right) -2\mbox{E}_E\left({\sqrt{2}}/{2}\right) \right)$ is a constant.
From these results, the force acting on the particle in the steady limit can be drawn and we obtain
\begin{equation}
\vec{f} = -6 \pi \mu a\left(\tens{I} + \frac{a}{\ell}\:
\tens{M} 
\right)\cdot
\vec{u}
\label{res_steady}
\end{equation}
where $\tens{M}$ is a diagonal mobility-like tensor whose components read as 
$$
M_{11}=M_{22}=\frac{1}{14}\mbox{E}_K\left(\frac{\sqrt{2}}{2}\right) \sim 0.1324 
\quad \mbox{and} \quad 
M_{33}=\frac{5}{14}\mbox{E}_K\left(\frac{\sqrt{2}}{2}\right) \sim 0.6622 \:.
$$
Note  that  though the off-diagonal components $M_{ij}$ ($i \neq j$) are found to be zero, as in the study by \cite{Chadwick74}, the perturbation force acting on the particle is not necessarily collinear with the particle velocity, since the three diagonal components are not identical. In other words,  a particle which is settling with an oblique trajectory in a vertically stratified fluid does experience a lift force.

\begin{figure}
\begin{center}
\begin{psfrags}
\psfrag{w}[c][b][1]{$\omega$}
\includegraphics[width=1.\linewidth]{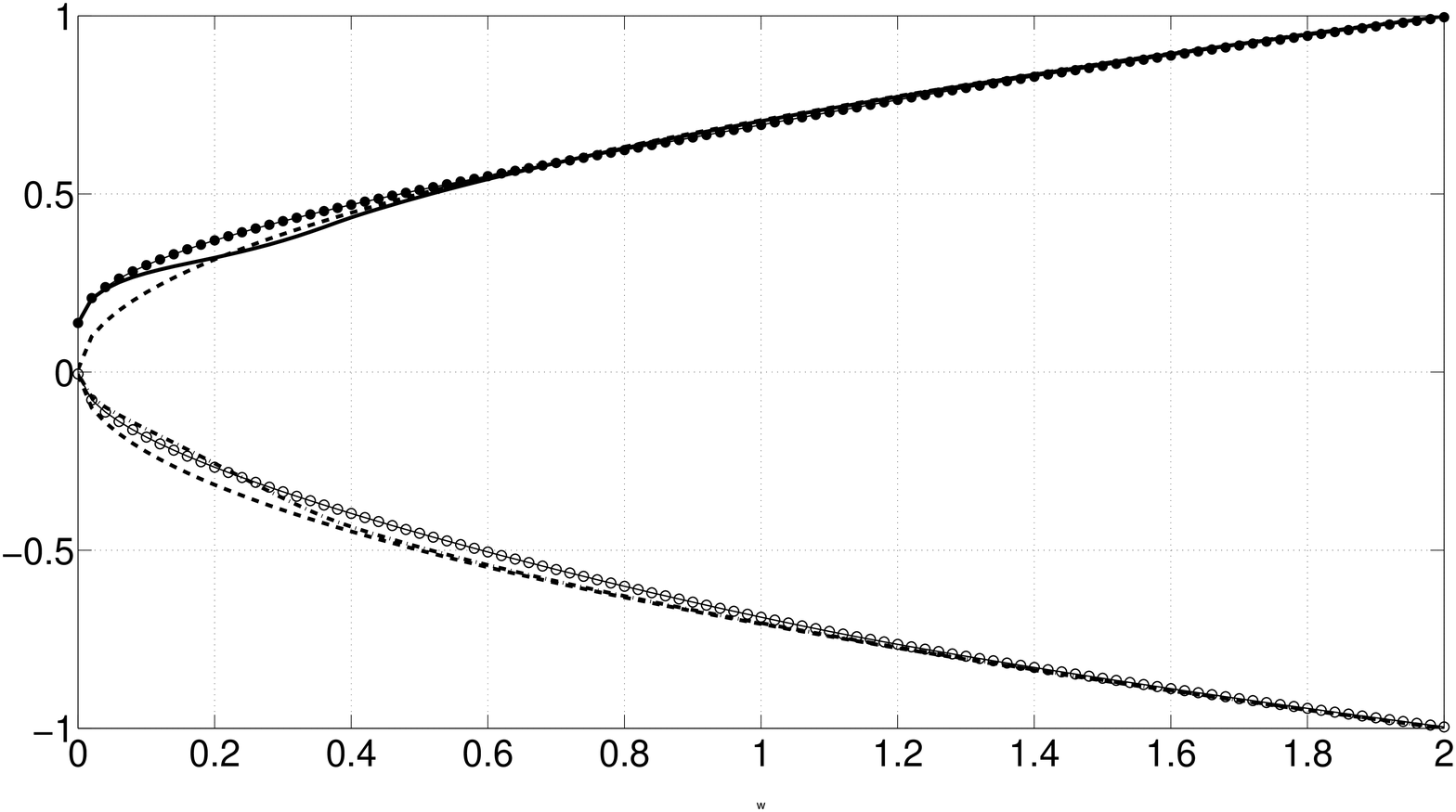}
\end{psfrags}
\caption{$h_\bot(\omega)$ versus $\omega$ for the cases $\mathrm{Pr}=7$ and $\mathrm{Pr}=0.7$. Continuous line: real part of $h_\bot$ ($\mathrm{Pr}=7$), dashed-dot line: imaginary part of $h_\bot$ ($\mathrm{Pr}=7$), symbols $\bullet$: real part of $h_\bot$ ($\mathrm{Pr}=0.7$), symbols $\circ$: imaginary part of $h_\bot$ ($\mathrm{Pr}=0.7$) and long-dashed line: $\pm (\sqrt{2}/{2})\sqrt{\omega}$ (history force in a homogeneous fluid)}
\label{figure2}
\end{center}
\end{figure}

\begin{figure}
\begin{center}
\begin{psfrags}
\psfrag{w}[c][t][1]{$\omega$}
\psfrag{pr1}[c][c][0.75]{$1/\sqrt{7}$}
\psfrag{pr2}[c][c][0.75]{$1/\sqrt{0.7}$}
\includegraphics[width=1.\linewidth]{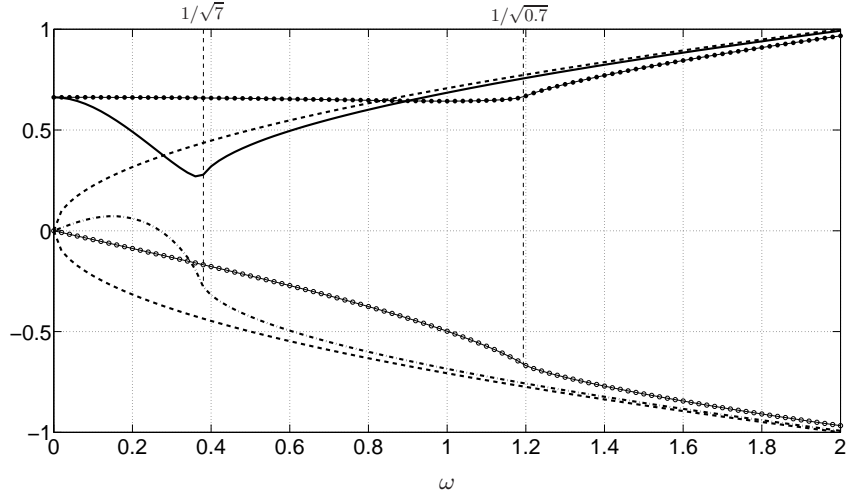}
\end{psfrags}
\caption{$h_\parallel(\omega)$ versus $\omega$ for the cases $\mathrm{Pr}=7$  and $\mathrm{Pr}=0.7$. Continuous line: real part of $h_\parallel$ ($\mathrm{Pr}=7$), dashed-dot line: imaginary part of $h_\parallel$ ($\mathrm{Pr}=7$), symbols $\bullet$: real part of $h_\parallel$ ($\mathrm{Pr}=0.7$), symbols $\circ$: imaginary part of $h_\parallel$ ($\mathrm{Pr}=0.7$) and long-dashed line: $\pm (\sqrt{2}/{2})\sqrt{\omega}$ (history force in a homogeneous fluid).}
\label{figure1}
\end{center}
\end{figure}

Apart from these asymptotic behaviours, the two components $h_\bot$ and $h_\parallel$ depend on the Prandtl number. As an illustration,  Figure 1 and 2 show the evolution of these components with respect to $\omega$, in the cases $\mathrm{Pr}=7$ and $\mathrm{Pr}=0.7$, which correspond, respectively, to the typical values of the Prandtl number in water (around 20 $^{\circ}$C) and in the air  (for a wide range of temperature). In both cases, it is seen, by comparisons with the force correction that would act on the particle in a homogeneous fluid, that the horizontal part of the force corrections, i.e. $h_\bot$, are only slightly affected by the stratification, in contrast with the vertical components $h_\parallel$. 

Another remarkable point to mention is that, in each case, the behaviour of the vertical force correction change abruptly for the  angular frequency corresponding to $1/\sqrt{\mathrm{Pr}}$ (see again figure \ref{figure1}). A more detailed analysis even shows that each of the three integral terms $I_4(\omega) + \overline{I_4}(-\omega)$, $I_5(\omega) + \overline{I_5}(-\omega)$ and $I_6(\omega) + \overline{I_6}(-\omega)$  actually admit two singularities (i.e. two poles)  located on the real axis and given by $\omega=\pm 1/\sqrt{\mathrm{Pr}}$, but the force correction remains, nevertheless, continuous since the different terms balance each other. Physically, the existence of such  singularities has to be related with an oscillating behaviour (in the temporal domain) of the drag force acting on the particle (given  that the Fourier transform of a sinusoidal term corresponds to  a delta function is the frequency domain). In particular, this phenomenon is certainly  linked to the establishing of the recirculating zones of the perturbation flow observed in the steady limit (see Ardekani \& Stoker 2010). In the temporal domain, the oscillating behaviour of the solution can also been inferred directly from equations (\ref{eq_phys_w}) and (\ref{eq_phys_rho}), at least qualitatively. Indeed, by considering a degenerate form of  these two equations, in which the diffusive (Laplacian) terms  and the pressure gradient term are neglected, we obtain (along $\vec{e}_3$, with $\vec{w}\cdot\vec{e}_3=w_3$) an equation of the form
$$
\frac{\partial^2 w_3}{ \partial t^2} + \frac{1}{\mathrm{Pr}} w_3 = 0\:, \quad
\mbox{whose solution reads as } \quad  w_3 \sim C_1 \cos\left(\frac{t}{\sqrt{\mathrm{Pr}}} \right) + C_2 \sin \left( \frac{t}{\sqrt{\mathrm{Pr}}} \right)\:.
$$
This approach, though based on simplified equations, allows the oscillating behaviour of  solution to be recovered. 
Finally, it is worth recalling that in this problem, the time has been normalised by $\ell^2/\nu$, so that in terms of dimensional variables,  the  two singular angular-frequencies correspond to $\pm N$, that is the Brunt-V\"ais\"al\"a frequency. It is physically sound to recover  this frequency which is generally  involved in stratified fluid disturbances. 

So far we have focused on the force in the frequency domain, which has been obtained by means of the matched-asymptotic-expansions method. Strictly speaking this method, which is based on a perturbation of the steady Stokes solution,  is restricted to perturbation flows which are only weakly unsteady, that is, in the present study, in the limit
$$
\omega \ll \frac{1}{\epsilon^2}\:.
$$
Fortunately, it turns out that the force given by (\ref{eq_res_1}) rapidly tends  to the classical Boussinesq-Basset history force (typically for $\omega \sim 2$), and in particular, before the limit $1/\epsilon^2$ is reached.  According to the fact that the Boussinesq-Basset force is expected to be recovered for highly unsteady  perturbation flows, since in these cases the unsteady term should dominates the buoyancy term in the fluid motion equation, we shall admit that the results obtained here are actually valid for any values of $\omega$.  This is an important point which will allow us to derive, in the following section, the force in the temporal domain by performing an inverse Fourier transform. Results obtained in the temporal domain will be illustrated through a simple example.

\subsection{force correction  in the temporal domain }
\vspace{11pt}

\begin{figure}
\begin{center}
\begin{psfrags}
\psfrag{t}[c][t][1]{$t$}
\includegraphics[width=1.\linewidth]{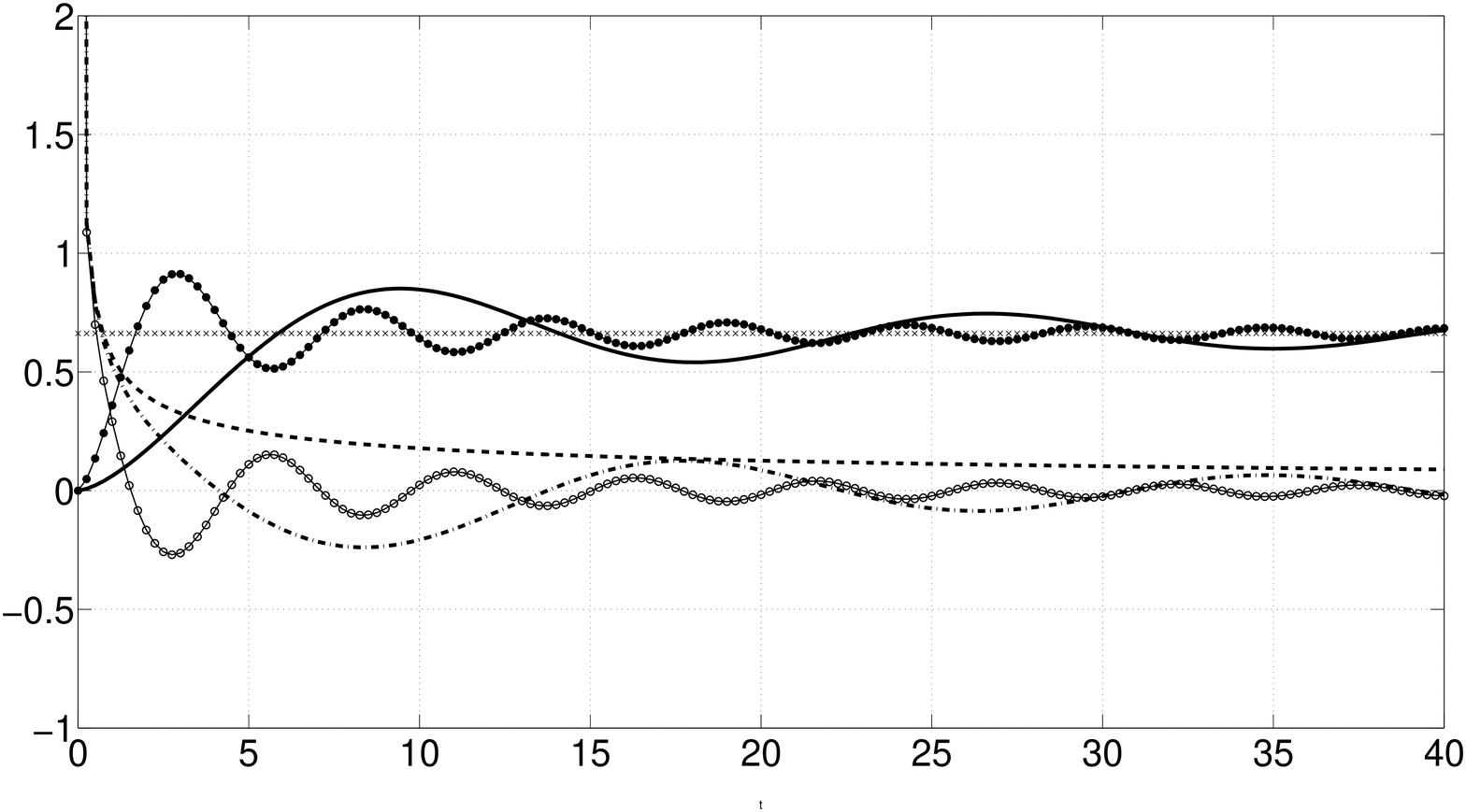}
\end{psfrags}
\caption{Evolutions of the two terms involved in  (\ref{response}) versus $t$ (non-dimensional time)  in response to an abrupt (vertical) motion of the particle, for the cases $\mathrm{Pr}=7$ and  $\mathrm{Pr}=0.7$. Continuous line: $\int_0^t \left(\tens{k}_1\right)_{33} \mbox{d} \tau$ ($\mathrm{Pr}=7$), dashed-dot line: $\left(\tens{k}_2\right)_{33}(t)$ ($\mathrm{Pr}=7$), symbols $\bullet$: $\int_0^t \left(\tens{k}_1\right)_{33} \mbox{d} \tau$ ($\mathrm{Pr}=0.7$), symbols $\circ$: $\left(\tens{k}_2\right)_{33}(t)$ ($\mathrm{Pr}=0.7$), long-dashed line: $1/\sqrt{\pi t}$ (history force kernel in a homogeneous fluid) and symbols $\times$: $({5}/{14})\mbox{E}_K\left({\sqrt{2}}/{2}\right) \sim 0.6622$.}
\label{figure3}
\end{center}
\end{figure}

\begin{figure}
\begin{center}
\begin{psfrags}
\psfrag{t}[c][c][1]{$t$}
\includegraphics[width=1.\linewidth]{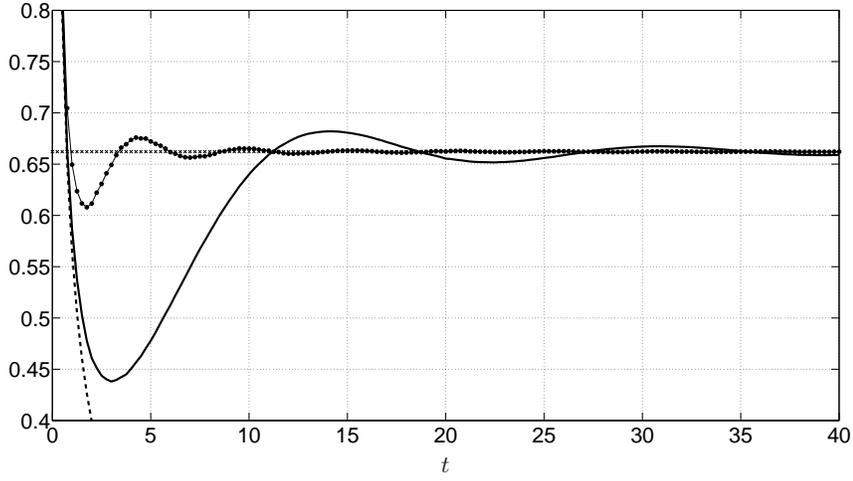}
\end{psfrags}
\caption{Vertical force correction versus $t$ (non-dimensional time) in response to an abrupt (vertical) motion of the particle, for the cases  $\mathrm{Pr}=7$ and $\mathrm{Pr}=0.7$. Continuous line: equation (\ref{response}) ($\mathrm{Pr}=7$), symbols $\bullet$: equation (\ref{response}) ($\mathrm{Pr}=0.7$),  long-dashed line: $1/\sqrt{\pi t}$ (history force kernel in a homogeneous fluid) and symbols $\times$: $({5}/{14})\mbox{E}_K\left({\sqrt{2}}/{2}\right) \sim 0.6622$.}
\label{figure4}
\end{center}
\end{figure}

For simplicity,  the  force is assumed here to be causal (e.g. the velocity of the particle is zero for $t<0$), and according to the form of (\ref{eq_h}) the calculation of the inverse (temporal) Fourier transform of 
(\ref{eq_res_1}) leads us to 
\begin{equation}
\mathcal{F}_t^{-1}\left(\frac{\vec{h}(\omega)}{8\pi^3}\right)  = \int_0^t \left(\tens{k}_1(t-\tau)\cdot  \vec{u}(\tau) +  \tens{k}_2(t-\tau)\cdot  \frac{ \mbox{d} \vec{u}}{ \mbox{d} \tau} \right)\mbox{d}\tau
\label{eq_res_temp}
\end{equation}
where $\tens{k}_1$ and $\tens{k}_2$ are two diagonal tensors whose components are given by
$$
\left(\tens{k}_1\right)_{11} = \left(\tens{k}_1\right)_{22} = 2\mathcal{R}\left(\mathcal{F}_t^{-1}(I_1) \right) \:, \quad 
\left(\tens{k}_1\right)_{33} =  2\mathcal{R}\left(\mathcal{F}_t^{-1}(I_4) \right)
$$
and 
$$
\left(\tens{k}_2\right)_{11} = \left(\tens{k}_2\right)_{22} = 2\mathcal{R}\left(\frac{\mbox{d}}{\mbox{d}t}\mathcal{F}_t^{-1}(I_2)  + \mathcal{F}_t^{-1}(I_3) \right) + \frac{3}{4\sqrt{\pi\:t}}\:,
$$
$$
 \left(\tens{k}_2\right)_{33}  = 2\mathcal{R}\left(\frac{\mbox{d}}{\mbox{d}t}\mathcal{F}_t^{-1}(I_5)  + \mathcal{F}_t^{-1}(I_6) \right)\:,
$$
where $\mathcal{R}(\cdot)$ denotes the real part of the complex functions. 

In practice, the inverse temporal Fourier transforms involved in the kernels  $\tens{k}_1$ and $\tens{k}_2$  have to be estimated numerically, and a particular attention must be paid to the calculation of the components $\left(\tens{k}_1\right)_{33} $ and $\left(\tens{k}_2\right)_{33}$ owing to the existence of the two poles mentioned previously (see discussion at the end of \S \ref{section_force_freq}). In order to overcome such a difficulty, we found it convenient to adopt the same technique as that used by Feynman to determine the causal retarded propagator in quantum mechanics. In brief, this method, which is based on the residue theorem, consists in adding a small positive imaginary part of the form $i \epsilon$ to  the two poles, which turns to be equivalent to perform a contour integration going clockwise over both poles (see for instance Appel 2002).  

In order to illustrate the force correction obtained in the temporal domain,  let us consider the behaviour of the force correction  in response to an abrupt change in velocity (along the vertical direction) modelled by the Heaviside unit step function 
$$
\vec{u} = H(t) \:\vec{e}_3
\quad \mbox{so that } \quad \frac{\mbox{d} \vec{u}}{\mbox{d}t} = \delta(t) \:\vec{e}_3\:,
$$
where $\delta(t)$ is the delta function. In this particular case,  the force correction reads as
\begin{equation}
\mathcal{F}_t^{-1}\left(\frac{\vec{h}(\omega)}{8\pi^3}\right)  = \left( \int_0^{t} \left(\tens{k}_1\right)_{33}(\tau) \: \mbox{d}\tau 
+ \left(\tens{k}_2\right)_{33} \right) \vec{e}_3\:,
\label{response}
\end{equation}
and similarly as in the previous section, results are drawn in the cases $\mathrm{Pr} = 7$ and $\mathrm{Pr}=0.7$ (i.e. typical values encountered in temperature-stratified water around 20 $^{\circ}$C, or in gas). Figure \ref{figure3} shows the distinct responses of the two terms involved in  (\ref{response}), and figure \ref{figure4} shows their sum. At short time, and in both cases, it is observed that 
$$
\int_0^{t} \left(\tens{k}_1\right)_{33}(\tau) \: \mbox{d}\tau  \sim \frac{1}{\sqrt{\pi t}} \quad \mbox{and } \quad
\left(\tens{k}_2\right)_{33}  \sim 0\:.
$$
This result, which is directly related to the asymptotic behaviour of the force correction  when $\omega \gg 1$, simply suggests that at the initial stage of the motion, the force correction is similar to that  which would be experienced by the particle in a homogeneous fluid.  
This result was expected, because at short time, the perturbation (i.e. the vorticity) generated by the sudden motion of the particle has not yet had time to diffuse to  the region where buoyancy effects alter the fluid flow. 
Figure \ref{figure4} also shows that the force corrections begin to separate from that corresponding to a homogeneous fluid  after a time of the order of a few units, which typically corresponds to the time the vorticity takes to diffuse  to the stratification length (i.e. $\ell^2/\nu$). According to the buoyancy effects, the rapid decrease of the forces obtained at short time is followed by a damped oscillation with a period given by $2\pi \sqrt{\mathrm{Pr}}$ (i.e. $T \sim 2\pi/N$ in dimensional variables). Note that the oscillations of the force corrections  are even more marked in figure \ref{figure3}  but, however, these terms evolve (almost) in anti-phase and (partially) balance each other. 
Evidently, at long time, the perturbation forces eventually tend to the constant provided in (\ref{res_steady}). 

\section{Concluding remarks}

This paper has investigated theoretically the hydrodynamic force acting on a particle undergoing an arbitrary time-dependent motion. This has been done by using the method of matching asymptotic expansions, which leads us to the results (\ref{eq_res_1}) in the frequency  domain, and to (\ref{eq_res_temp})  in the temporal domain. The theory presented is valid, provided that the condition (\ref{condition2}) is satisfied, and also that the Reynolds and the Péclet numbers involved in these relations remain small compared to unity.

 To give a first example, these results can be applied to predict 
the motion of large light particles (like porous aggregates)  in highly temperature-stratified atmospheres, such as those encountered in 
fire engineering, for instance. In these cases, according to the perfect gas law,  the Brunt-V\"ais\"ala frequency, which can be approximated by 
$$
N \sim \sqrt{\frac{1}{T_\infty} \frac{\mbox{d}T}{\mbox{d}z} g }\:,
$$
may reach values up to the order of unity (s$^{-1}$), therefore leading us to a stratification length of a few millimetres.  In these environments,
the force correction obtained in this investigation may be useful to predict the motion of  a particle smaller than $\ell$, but not too much, i.e. typically ranging from a few hundred $\mu$m to approximately 1 or 2 mm,  
provided that its velocity remains small enough to ensure  the smallness of the Reynolds number (typically for $u$ ranging from 0.01 to 0.1 m/s in the case $\mathrm{Re} \sim 0.1$). Note that in temperature-stratified gas, the Prandtl number value remains around 0.7 even for high temperatures, so that 
according to the classical relation $\mathrm{Pe} = \mathrm{Re}\: \mathrm{Pr}$, the Péclet number is slightly smaller than the Reynolds number in (\ref{condition2}).  

A second example which is interesting to consider, and where the correction force may be influential in determining the trajectories of particles, is temperature-stratified water. In natural environment, and as stated in the previous section, the typical value of the  Prandtl number of temperature-stratified water is around 7 (at 20 $^{\scriptsize{0}}$ C), and that of $\gamma$ is about 1 kg/m$^{-4}$, so that the stratification length is found to be around $\ell \sim 1.9$ mm. Again, corrections should be applied for particles with a radius of a few dozen $\mu$m. In this case, satisfying the  condition $1/\mbox{Pe} \gtrsim \ell/a$, as required by (\ref{condition2}), provides us with a maximal value of the velocity of the particle $u \lesssim {\kappa}/{\ell}$
for which the theory remains valid. Typically, in temperature-stratified water,  the ratio $\kappa/\ell$ is approximately around $10^{-4}$ (m/s), so that the particle velocity should not exceed a few radius per second, except if its motion is highly unsteady. 
Let us also mention that the theoretical results of the present study have been derived  after assuming that the viscosity of the fluid is constant, at least in the vicinity of the inclusion, i.e. in a region corresponding to $r \lesssim  \ell$.  In the two examples discussed previously, the variations of the viscosity due to inherent stratification of the fluid can be estimated by evaluating the term $\ell ({\mbox{d}\nu}/{\mbox{d}z})$. 
In both cases, it is found that these variations are indeed negligible in comparison with the viscosity evaluated at the initial location of the particle, since 
$$
\frac{\ell}{\nu} \:\displaystyle{\frac{\mbox{d}\nu}{\mbox{d}z}}\sim O(10^{-4})\:.
$$
However, though the variations of the viscosity can be neglected in the vicinity of the particle, the validity of the theoretical results can be called into question if the particle is followed over a sufficiently large distance (in the vertical  direction), for the variations of viscosity to become significant.

As a final remark, and  as already mentioned in \S2,  in the study by \cite{Zvirin75}, the force correction obtained for a particle settling in a slightly diffusing stratified fluid (i.e. at larger Péclet numbers than in the present paper) has been found to depend on a stratification number defined in (\ref{def_Ri}). 
In  the paper by \cite{Yick09}, it turns out that a part of their numerical runs (i.e. those in which $\mathrm{Re}=0.05$ and $\mathrm{Pr}=7$) 
actually  partially matches the conditions required in this investigation, at least for the smallest values of $\mathrm{Ri}$. According to the value of the Péclet number corresponding to these numerical simulations (i.e. $\mathrm{Pe}= 0.35$) the quasi-steady force  provided by (\ref{res_steady})  reads as
$$
\vec{f} = 6 \pi \left(1 +0.6622\:(0.35)^{1/4}  \:\mathrm{Ri}^{1/4}\right)\:, \quad \mbox{where}  \quad 0.6622 \:(0.35)^{1/4} \sim  0.5093 
$$
and  seems to be in quite good agreement with the results presented by the authors, although it is difficult to make an accurate comparison.

\end{document}